\documentclass[showpacs,amsmath,nofootinbib,amssymb, epsfig]{revtex4}
\usepackage{graphicx} 
\usepackage{dcolumn}
\usepackage{bm}

\usepackage{color}

\usepackage{amssymb}

\begin{document}

\title{ HOLOGRAPHIC DARK ENERGY MODELS WITH INTERACTIONS AND ENTROPY CORRECTIONS FOR DIFFERENT CUTOFFS IN THE BRANS–DICKE COSMOLOGY}
\author{F. Darabi}\email{f.darabi@azaruniv.ac.ir} \author{F. Felegary}\email{falegari@azaruniv.ac.ir}
\affiliation{Department of Physics, Azarbaijan Shahid Madani University, Tabriz, 53714-161 Iran}

\date{\today}

\begin{abstract}
In the framework of Brans-Dicke cosmology, we have studied the interacting power-law and logarithmic entropy-corrected holographic dark energy models with different cut-offs. The Brans-Dicke parameter is investigated versus
the conditions for the acceleration and phantom phases to show that which entropy corrected model can exhibit acceleration with or without the phantom phase at early and present time universes. Moreover, the classical stability or instability of the interacting power-law and logarithmic entropy-corrected holographic dark energy models with different cut-offs is determined at early and present time.
\end{abstract}
\vspace{1cm}
\maketitle

\section{Introduction}
Recent cosmological and astrophysical data gathered from the observations type Ia supernovae (SNeIa), cosmic microwave background radiation (CMB) and large scale structure (LSS)  convincingly suggest that the observable universe experiences an accelerated expansion phase \cite{Riess} that this expansion may be driven by a mysterious energy component with negative pressure so called dark energy. The cosmological constant is the simplest candidate for dark energy, called $\Lambda$CDM model. Nowadays cosmologists know that the cosmological constant suffers from two difficulties, the fine- tuning and the cosmic coincidence problems \cite{Copeland,Weinberg}. The cosmic coincidence problem needs that universe behaves in such as a form that the ratio of dark matter to dark energy densities must be constant of order unity\cite{Zimdahl}. There are various models for dark energy including the cosmological constant, Quintessence, K-essence, Phantom, Quintom, Chaplygin gas, Thachyon and modified gravity \cite{Padmanabhan,Copeland}. Recently, the holographic dark energy (HDE) as a new dark energy model based on the holographic principle was suggested \cite{Susskind}
\begin{equation}
\rho_{\Lambda}=3c^{2}M_{P}^{2}L^{-2},
\end{equation}
where c is a numerical constant, $L$ is the cut-off radius and $M_{P}$ is the reduced Planck mass. In the holographic dark energy model the Bekenstein-Hawking
entropy relation
$S_{BH}=A/4G$ plays an essential role and is satisfied on the horizon \cite{Wald} where $A\sim L^{2}$ is the area of horizon. Since this model is closely connected to the area law of entropy, hence any correction to this law will affect the energy density of the holographic dark energy model. These corrections
 may arise due to the quantum field theory  \cite{Radicella,Jamil},  thermal and quantum fluctuations in LQG \cite{Ashtekar,Rovelli,Ghosh,Medved, Meissner},
 and string theory \cite{Wald1}.

One correction to the  entropy is the power-law correction   \cite{Radicella}
\begin{equation}
S=A/4G(1-K_{\alpha}A^{1-\frac{\alpha}{2}}),
\end{equation}
where $\alpha$ is a dimensionless constant and
\begin{equation}
K_{\alpha}=\frac{\alpha(4\pi)^{\frac{\alpha}{2}-1}}{(4-\alpha)r_{c}^{2-\alpha}},
\end{equation}
where $r_{c}$ is the cross over scale. This correction results in a modification of holographic dark energy called power-law entropy corrected holographic dark energy (PLECHDE) as \cite{Sheykhi}
\begin{equation}
\rho_{\Lambda}=3c^{2}M_{P}^{2}L^{-2}-\beta M_{P}^{2}L^{-\alpha}, \label{rho1}
\end{equation}
where $\alpha$ and $\beta$ are two constants of the order of  unity. In Ref.\cite{Radicella},
it was demonstrated that the generalized second law of thermodynamics for the universe with the power-law corrected entropy is satisfied for $\alpha>2$. For $\alpha>2$ the second term in Eq.(\ref{rho1}) is comparable to the first term when $L$ takes a very small value, thus the correction has a physical meaning only at early universe and  is ignorable when the universe becomes large \cite{Sheykhi}.

 Another corrected entropy takes on the following form \cite{Majhi1}
\begin{equation}\label{E-C}
S_{BH}=\frac{A}{4G}+\tilde{\alpha}\ln(\frac{A}{4G})+\tilde{\beta},
\end{equation}
where $\tilde{\alpha}$ and $\tilde{\beta}$ are two dimensionless constants.  The logarithmic term also appears in a model of entropic cosmology, capable of unifying the early time inflation and late-time acceleration of the universe \cite{Cai}. The entropy-area relation can be expanded in a series of infinite terms, however the contribution  of extra terms are negligible due to the smallness of reduced Planck constant $\hbar$. Hence, the most important leading term in the expansion is left as the logarithmic one, which has been  considered in this paper.  Considering the corrected entropy-area relation (\ref{E-C}) and following the derivation of HDE, the corresponding energy density  will be modified. Wei, has recently proposed the energy density $\rho_{\Lambda}$ of the  logarithmic entropy-corrected holographic dark energy (LECHDE) in the following form \cite{Wei}
\begin{equation}
\rho_{\Lambda}=\frac{3c^{2}\phi^{2}}{4\omega L^{2}}+\frac{\alpha}{L^{4}}\ln(\frac{L^{2}\phi^{2}}{4\omega})+\frac{\beta}{L^{4}},\label{rho}
\end{equation}
where $\alpha$ and $\beta$ are dimensionless constants. The second and third terms in Eq.(\ref{rho}) are comparable to the first term only when $L$ takes a very small value, hence the corrections given by these  terms  have a physical meaning only at early  universe. When the universe becomes large, these corrections  are ignorable and
LECHDE reduces to the ordinary HDE. Therefore, in the following first we will investigate the
state of the universe at present time and then study the effect of different
cut-offs, considering the corrections at early time, on the state of the universe \cite{Sheykhi1}. Since HDE density corresponds to a dynamical cosmological constant, we need a dynamical frame as alternative theory of general relativity to accommodate HDE density.

 The scalar-tensor theory was first established as an alternative to general
 relativity \cite{Jordan,Brans}. Then, it played the essential role
in solving the main problems of standard cosmology in the context of inflationary
scenario. The main motivation to use the scalar field models in quest for solving the recent DE problem in cosmology lies in the particle physics as well as string theory. In order to solve the DE problem, several
models of dynamical dark energy with time evolving equation
of state have been proposed in the context of scalar field models such as quintessence \cite{qui}, k-essence \cite{ess}, phantom \cite{pha},
tachyon \cite{tac}, and quintom \cite{quin}.  Reconstruction of the holographic  and agegraphic dark energy models with scalar fields is also one of the most attractive subjects  studied in this direction \cite{Almeida,Granda,Zhang, karami}. The investigation on the holographic dark energy  model in the framework of Brans-Dicke (BD) theory is one of these attempts with interesting results
which, for example, implies that one can not generate phantom-like equation of state from a holographic dark energy model in non-flat universe in the framework of Brans-Dicke cosmology \cite{Setare}.

In this work, we aim to generalize the study in \cite{Setare} and consider the interacting power-law and logarithmic entropy-corrected  holographic dark energy model in Brans-Dicke cosmology and obtain the equations of state parameter  for different viable cut-offs. We also consider the correspondence between the  interacting power-law and logarithmic entropy-corrected holographic dark energy models with different cut-offs in Brans-Dicke cosmology in one
hand, and quintessence and tachyon scalar field models in non-flat universe,
on the other hand. Moreover, we  perform the stability analysis of the models by determining the squared sound speed, $v_{s}^{2}=\frac{dp}{d\rho}$ \cite{P1}.
If $v_{s}^{2}<0$,  we have the classical instability against given perturbation. In contrast $v_{s}^{2}>0$ may lead to a stable universe against perturbations.
\section{Interacting  entropy-corrected  holographic dark energy model in Brans-Dicke cosmology}
The action of Brans-Dicke theory in the canonical form may be written as \cite{Arik}
\begin{equation}
\int d^4 x \sqrt{g}(-\frac{1}{8\omega}\phi^2 R + \frac{1}{2}g^{\mu \nu} \partial_{\mu}\phi \partial_{\nu}\phi + L_{M}),
\end{equation}
where $g$ is the determinant of the tensor metric $g^{\mu\nu}$, $\omega$ is the Brans-Dicke parameter, $R$ is the Ricci scalar curvature and $L_{M}$ is the lagrangian of the matter. Variation of the action with respect to the metric $g^{\mu \nu}$ and the Brans-Dicke scalar field $\phi$ obtains
\begin{eqnarray}
\phi G_{\mu \nu}=-8\pi T^{M}_{\mu\nu}-\frac{\omega}{\phi}(\phi_{, \mu}\phi_{,\nu}-\frac{1}{2}g_{\mu \nu} \phi_{, \lambda} \phi_{, \lambda} 
-\phi_{;\mu;\nu}+\Box\phi g_{\mu\nu}),\label{G}
\end{eqnarray}
\begin{eqnarray}
\Box \phi = \frac{8\pi}{2\omega+3} T^{M \lambda}_{\lambda}.\label{box}
\end{eqnarray}
Here, $T^{M}_{\mu \nu}$ is the energy-momentum tensor of the matter fields.
The Friedman-Robertson-Walker (FRW) universe is defined by
\begin{equation}
ds^{2}= dt^{2}-a(t)^{2}(\frac{dr^{2}}{1-kr^{2}}+r^{2}d\Omega^{2}),\label{frw}
\end{equation}
where $a(t)$ and $k$ are scale factor and curvature parameter, respectively. Using Eq.(\ref{frw}), the field equations (\ref{G}) and (\ref{box}) are simplified to
\begin{equation}
\frac{3}{4\omega}\phi^{2}(H^{2}+\frac{k}{a^{2}})-\frac{1}{2}\dot{\phi}^{2}+\frac{3}{2\omega}H\dot{\phi}\phi
=\rho_{m}+\rho_{\Lambda},\label{frid}
\end{equation}
\begin{eqnarray}
-\frac{1}{4\omega}\phi^{2}(2\frac{\ddot{a}}{a}+H^{2}+\frac{k}{a^{2}})-\frac{1}{\omega}H\dot{\phi}\phi\nonumber\\
-\frac{1}{2\omega}\ddot{\phi}\phi
-\frac{1}{2}(1+\frac{1}{\omega})\dot{\phi}^{2}=p_{\Lambda},\label{Plambda}
\end{eqnarray}
\begin{equation}
\ddot{\phi}+3H\dot{\phi}-\frac{3}{2\omega}(\frac{\ddot{a}}{a}+H^{2}+\frac{k}{a^{2}})\phi=0,
\end{equation}
where $H={\dot{a}}/{a}$, $\rho_{m}$, $\rho_{\Lambda}$ and $p_{\Lambda}$ are the Hubble parameter, the pressureless matter density, the energy density and pressure of dark energy, respectively. We suppose that there is an interaction between entropy-corrected  holographic model of dark energy and dark matter as follows
\begin{equation}
\dot{\rho_\Lambda}+3H(1+\omega_\Lambda)\rho_{\Lambda}=-Q,\label{dot rho}
\end{equation}
\begin{equation}
\dot{\rho_M}+3H\rho_M=Q,\label{QM}
\end{equation}
where $Q=3b^{2}H(\rho_{\Lambda}+\rho_{m})$ is the interaction term and $b^{2}$ is a coupling constant.
Moreover, following \cite{J} we  suppose  that the BD field behaves as $\phi=a^{n}$
so that
\begin{equation}
\dot{\phi}=nH\phi~~~~,~~~~\ddot{\phi}=(n^{2}H^{2}+n\dot{H})\phi.\label{scale}
\end{equation}
A case of specific interest is when $n$ is small while $\omega$ is large, so that the product $n\omega$ results in a value of order unity \cite {J}.
The fractional energy densities are given by
\begin{equation}
\Omega_{M}=\frac{4\omega\rho_{m}}{3\phi^{2}H^{2}},
\end{equation}
\begin{equation}
\Omega_{k}=\frac{k}{a^{2}H^{2}},\label{k}
\end{equation}
\begin{equation}
\Omega_{\Lambda}=\frac{4\omega\rho_{\Lambda}}{3\phi^{2}H^{2}}.\label{Lambda}
\end{equation}
Taking time derivative of (\ref{frid}) gives
\begin{equation}
H(z)=H_{0}\Big[\frac{\Omega_{M_{0}}(1+z)^{2n+3}}{1+\Omega_{k}+\frac{2}{3}n(3-n\omega)-\Omega_{\Lambda}}\Big]^\frac{1}{2}\label{Hz},
\end{equation}
where we have converted time derivative to a derivative with respect to $z$,  and zero-indices quantities indicates the 
 present
time values. 
Also, we obtain the deceleration parameter as
\begin{equation}
q=-\frac{\ddot{a}a}{\dot{a}^{2}}=-1-\frac{\dot{H}}{H^{2}}.\label{deceleration}
\end{equation}
Taking time derivative of (\ref{k}) yields
\begin{equation}
\frac{d \Omega_{k}}{dz}=2\Omega_{k} \Big((1+z)^{-1}-\frac{\frac{dH}{dz}}{H} \Big),\label{kk}
\end{equation}
where we have used using $\frac{d\Omega_{k}}{dt}=-H(1+z)\frac{d\Omega_{k}}{dz}$
and  $\frac{dH}{dt}=-H(1+z)\frac{dH}{dz}$. On the other hand, taking time derivative of (\ref{Lambda}), using (\ref{dot rho}) and $\frac{d\Omega_{\Lambda}}{dt}=-H(1+z)\frac{d\Omega_{\Lambda}}{dz}$, yields
\begin{eqnarray}
\frac{d\Omega_{\Lambda}}{dz}&=&(1+z)^{-1}\Big[3b^{2}(\Omega_{\Lambda}+\Omega_{m})+3(1+\omega_{\Lambda})\Omega_{\Lambda}
+2n\Omega_{\Lambda}-2\Omega_{\Lambda}(1+z)\frac{\frac{dH}{dz}}{H} \Big]. \label{o}
\end{eqnarray}
Taking time derivative of (\ref{Hz}),  using Eqs.(\ref{kk}), (\ref{o}) and  $\frac{dH}{dt}=-H(1+z)\frac{dH}{dz}$, yields
\begin{eqnarray}
\frac{\frac{dH}{dz}}{H}=-\frac{(2n+3+3b^{2})(1+\Omega_{k}-\frac{2}{3}n^{2}\omega+2n)}{2(1+z)(-1+\frac{2}{3}n^{2}\omega-2n)}
-\frac{-2\Omega_{k}+3\Omega_{\Lambda}\omega_{\Lambda}}
{2(1+z)(-1+\frac{2}{3}n^{2}\omega-2n)}. \label{dH}
\end{eqnarray}
Therefore, using Eq.(\ref{dH}), the deceleration parameter is obtained
\begin{eqnarray}
q=-1-\frac{(2n+3+3b^{2})(1+\Omega_{k}-\frac{2}{3}n^{2}\omega+2n)-2\Omega_{k}+3\Omega_{\Lambda}\omega_{\Lambda}}{2(-1+\frac{2}{3}n^{2}\omega-2n)}
.
\end{eqnarray}
Moreover , using Eqs.(\ref{rho1}), (\ref{rho}) and (\ref{Lambda}), we can
obtain the following relations
\begin{equation}
\beta=\frac{3c^{2}}{L^{-\alpha+2}}\Big(1-\frac{L^{2}H^{2}\Omega_{\Lambda}}{c^{2}}\Big),\label{beta1}
\end{equation}
\begin{equation}
\alpha \ln(\frac{\phi^{2}L^{2}}{4\omega})+\beta=L^{4}\Big(H^{2}\Omega_{\Lambda}-\frac{c^{2}}{
L^{2}}\Big)\frac{3\phi^{2}}{4\omega},\label{beta11}
\end{equation}
which will be used in the derivation of $\omega_{\Lambda}$ in the following.
For power-law and logarithmic entropy-corrected holographic  dark energy models with any cut-off in Brans-Dicke cosmology, taking time derivative of (\ref{rho1}) and (\ref{rho}), using (\ref{dot rho}), and using (\ref{beta1}),
(\ref{beta11}) we obtain,
respectively
\begin{eqnarray}
\omega_{\Lambda}=-1-\frac{b^{2}}{\Omega_{\Lambda}}\Big(1+\Omega_{k}-\frac{2}{3}n^{2}\omega+2n\Big)
-\frac{2c^{2}}{L^{2}H\Omega_{\Lambda}}\Big[(1+z)\frac{\frac{dL}{dz}}{L}\Big(1-\frac{\alpha}{2}(1-\frac{L^2H^2\Omega_{\Lambda}}{c^2}) \Big)+
\frac{L^2H^2\Omega_{\Lambda}n}{c^2} \Big], \label{Omega1}
\end{eqnarray}
\begin{eqnarray}
\omega_{\Lambda}=-1-\frac{b^{2}}{\Omega_{\Lambda}}(1+\Omega_{k}-\frac{2}{3}n^{2}\omega+2n)
-
(n-\frac{\frac{dL}{dz}}{L}(1+z))\Big[\frac{2c^{2}}{3L^{2}H^2\Omega_{\Lambda}}
+\frac{8\alpha \omega (1+z)^{2n}}{9L^{4}H^2\Omega_{\Lambda}} \Big]
-\frac{4}{3}(1+z)\frac{\frac{dL}{dz}}{L}.
\label{Omega}
\end{eqnarray}

One of the most important quantities for cosmological evolution of the universe is the squared sound speed, $v_{s}^{2}=\frac{dp}{d\rho}$.
The squared sound speed of dark energy fluid is given by
\begin{equation}
v_{s}^{2}=\frac{dp_{\Lambda}}{d\rho_{\Lambda}}=\frac{\dot{p_{\Lambda}}}{\dot{\rho_{\Lambda}}}.
\end{equation}
Also, we have
\begin{equation}
v_{s}^{2}=\omega_{\Lambda}-\frac{\dot{\omega}_{\Lambda}}{3H(1+\omega_{\Lambda}^{eff})}.\label{vss}
\end{equation}
Here $\omega_{\Lambda}^{eff}=\omega_{\Lambda}+\frac{Q}{3H\rho_{\Lambda}}$ corresponds to the effective equation of state parameter for dark energy.
For $v_{s}^{2}>0$, there is a classical stability and for  $v_{s}^{2}<0$, there is a classical instability \cite{kim1}. 
\section{Interacting  entropy-corrected  holographic dark energy model with Hubble cut-off}
The Hubble cut-off is considered as \textcolor[rgb]{1,0,0.501961}{}
\begin{equation}
L=H^{-1}.\label{hubble}
\end{equation}
 
\subsection{PLECHDE model}

 Using Eqs.(\ref{Hz}),
 (\ref{dH}), and (\ref{Omega1}), we obtain
\begin{eqnarray}
\omega_{\Lambda}=  \Big\{-1-\frac{b^{2}}{\Omega_{\Lambda}}(1+\Omega_{k}-\frac{2}{3}n^{2}\omega+2n)
-\frac{2c^{2}H_{0}}{\Omega_{\Lambda}}\Big[\frac{\Omega_{M_{0}}(1+z)^{2n+3}}{1+\Omega_{k}+\frac{2}{3}n(3-n\omega)-\Omega_{\Lambda}} \Big]^{\frac{1}{2}}\times
~~~~~~~~~~~~~~~~~~~~~~~~\nonumber\\
\Big[\Big(\frac{(2n+3+3b^{2})(1+\Omega_{k}-\frac{2}{3}n^{2}\omega+2n)-2\Omega_{k}}{2(-1-2n+\frac{2}{3}n^{2}\omega)} \Big)\times
\Big(1-\frac{\alpha}{2}
(1-\frac{\Omega_{\Lambda}}{c^{2}}) \Big)+\frac{n\Omega_{\Lambda}}{c^{2}}\Big] \Big\}\times~~~~~~~~~~~~~~~~~~~~~ \nonumber\\
\Big\{(-1-2n+\frac{2}{3}n^{2}\omega)\times~~~~~~~~~~~~~~~~~~~~~~~~~~~~~~~~~~~~~~~~~~~~~~~~~~~~~~~~~~~~~~~~~~~~~~~~~~~~~~~~~~~~~~~~~~~~~~~~~~~\nonumber\\\Big[-1-2n+\frac{2}{3}n^{2}\omega+3c^{2}H_{0}\Big(1-\frac{\alpha}{2}
(1-\frac{\Omega_{\Lambda}}{c^{2}}) \Big)\times
\Big[\frac{\Omega_{M_{0}}(1+z)^{2n+3}}{1+\Omega_{k}+\frac{2}{3}n(3-n\omega)-\Omega_{\Lambda}} \Big]^{\frac{1}{2}}\Big]^{-1} \Big\}.~~~~~~~~~~~~~~~~~~~
\end{eqnarray}\label{wh}

For present time, we consider $z\simeq0$, $\alpha\simeq 0$, $c^{2}\simeq1.1$ \cite{ma}, $n\simeq0.005$ \cite{Lu}, $H_{0}\simeq67.8$, $\Omega_{\Lambda}\simeq0.7$ $\Omega_{M_{0}}\simeq0.27$, $\Omega_{k}\simeq0$ and $b^{2}=0.02$ \cite{ma}. The condition for acceleration as $q<0$, results
in $-10000\lesssim\omega\lesssim10000$ which is almost the same range for
having the phantom phase $\omega_{\Lambda}<-1$.

At early universe, using Eqs.\eqref{deceleration}, (\ref{Omega1}) and $H\simeq const$
\cite{Shey}, we obtain $q\simeq-1$ and
\begin{equation}
\omega_{\Lambda}=-1-2nH.\label{wh1}
\end{equation}
For early time, the condition $\omega_{\Lambda}<-1$ requires  $n>0$, so $n\simeq0.005$ \cite{Lu} is consistent with the phantom phase. Therefore, the PLECHDE model with Hubble cut-off  and $-10000\lesssim\omega\lesssim10000$
supports the inflationary and accelerating universe with phantom phase at early and present universe, respectively.  

Using  $\omega_{\Lambda}$ and its time derivative corresponding to  early and present times, and using Eq.(\ref{vss}) with $b^{2}=0, 0.02, 0.04, 0.06$,  we get $v_{s}^{2}<0$ and $v_{s}^{2}>0$ respectively,
indicating that PLECHDE model has classical instability at early time and classical stability at present time. 

\subsection{LECHDE model}
For LECHDE model, using Eqs.(\ref{Hz}),  (\ref{dH}), and (\ref {Omega}) we obtain
\begin{eqnarray}
\omega_{\Lambda}=
\Big[-1-n[\frac{2c^{2}}{3\Omega_{\Lambda}}+\frac{8\alpha \omega H_0^{2}\Omega_{M_{0}}(1+z)^{4n+3}}{9\Omega_{\Lambda}(1+\Omega_{k}
+\frac{2n}{3}(3-n\omega)-\Omega_{\Lambda})}] 
+[\frac{-2\Omega_{k}+(2n+3+3b^2)(1+\Omega_{k}
-\frac{2n^{2}\omega}{3}+2n)}{2(-1+\frac{2}{3}n^2 \omega-2n)}]~~~\nonumber \\ 
\times[\frac{2c^{2}}{3\Omega_{\Lambda}}
+\frac{8\alpha \omega H_0^{2}\Omega_{M_{0}}(1+z)^{4n+3}}{9\Omega_{\Lambda}(1+\Omega_{k}
+\frac{2n}{3}(3-n\omega)-\Omega_{\Lambda})}-\frac{4}{3}]
-\frac{b^{2}}{\Omega_{\Lambda}}(1+\Omega_{k}-\frac{2}{3}n^2 \omega+2n)\Big]\times~~~~~~~~~~~~~~~~~~~~~~~~~~~ \\ \nonumber
\Big[1-(\frac{3\Omega_{\Lambda}}{2(-1+\frac{2}{3}n^{2}\omega-2n)})\times [\frac{2c^{2}}{3\Omega_{\Lambda}}+\frac{8\alpha \omega H_0^{2}\Omega_{M_{0}}(1+z)^{4n+3}}{9\Omega_{\Lambda}(1+\Omega_{k}
+\frac{2n}{3}(3-n\omega)-\Omega_{\Lambda})} \nonumber
-\frac{4}{3}]\Big]^{-1}.\\ \nonumber~~~~~~~~~~~~~~~~~~~~
\label{w2q}
\end{eqnarray}
 With the same numerical set up mentioned before, in order to have $q<0$  at present time, we obtain $\omega\lesssim8383$. However, $\omega_{\Lambda}<-1$
requires $\omega\lesssim-2860$.
In other words, $\omega\lesssim-2860$ supports the phantom phase as well
as an accelerating universe. For $-2860\lesssim\omega\lesssim8380$, we have
an accelerating universe without the phantom phase.  

At early universe, using Eqs.\eqref{deceleration}, (\ref{Omega}), and $H\simeq const$, we obtain $q\simeq-1$ and
\begin{eqnarray}
\omega_{\Lambda}=-1-n[\frac{2c^{2}}{3\Omega_{\Lambda}}+\frac{8\alpha \omega H^{2}(1+z)^{2n}}{9\Omega_{\Lambda}}].
\label{wq1}
\end{eqnarray}
Using the parameter values $z\simeq 10^4$, $\alpha\simeq 1$, $c^{2}\simeq1.1$, $\Omega_{\Lambda}\simeq1$, the condition $\omega_{\Lambda}<-1$ or $\omega_{\Lambda}>-1$ requires  $\omega\gtrsim-H^{-2}$ or $\omega\lesssim-H^{-2}$, respectively .

     Therefore,  LECHDE model with Hubble cut-off  and $\omega\lesssim-2860$
supports the inflationary and accelerating universe with phantom phase at  present universe; however, at early universe we may have inflationary universe
with or without phantom phase corresponding to $\omega\gtrsim-H^{-2}$ or $\omega\lesssim-H^{-2}$, respectively.

 Using the same procedure as that of PLECHDE model, for LECHDE model we obtain $v_{s}^{2}<0$ for both early and present times which accounts for the classical instability at both eras.

{\section{Interacting  entropy-corrected  holographic dark energy model with apparent horizon cut-off}
The apparent horizon cut-off is given as 
\begin{equation}
L=\tilde{r_{A}}=\frac{1}{\sqrt{(H^{2}+\frac{k}{a^{2}})}}. \label{apparent}
\end{equation}

\subsection{PLECHDE model}

For power-law entropy-corrected holographic  dark energy model,
using Eqs. (\ref{Hz}),  (\ref{dH}), (\ref{Omega1}) and (\ref{apparent}), we can obtain EOS parameter at the present time
\begin{eqnarray}
\omega_{\Lambda}=\Big\{-1-\frac{b^{2}}{\Omega_{\Lambda}}(1+\Omega_{k}-\frac{2}{3}n^{2}\omega+2n)
-\frac{2c^{2}H_{0}(1+\Omega_{k})}{\Omega_{\Lambda}}\Big[\frac{\Omega_{M_{0}}(1+z)^{2n+3}}{1+\Omega_{k}+\frac{2}{3}n(3-n\omega)-\Omega_{\Lambda}} \Big]^{\frac{1}{2}}\times~~~~~~~~~~~~~~~~~~~~~~~~~~~~~~~~~~~~~~~~~~~~~~~~~~~~~~~~~~~~~~~~~~~~~~~~~~~~~~~~~~~~~~~~~~~~~~~~~~~~~~~~~~~~~~~
~~~~~~~~~~~~~~~~~~~~~~~~~\nonumber\\
\Big[\Big((2n+3+3b^{2})(1+\Omega_{k}-\frac{2}{3}n^{2}\omega+2n)
-2\Omega_{k}(-2n+\frac{2}{3}n^{2}\omega)\Big)\times
\Big(2(1+\Omega_{k})(-1-2n+\frac{2}{3}n^{2}\omega)\Big)^{-1}~~~~~~~~~~~~~~~~~~~~~~~~~~~~~~~~~~~~~~~~~~~~~~~~~~~~~~~~~
~~~~~~~~~~~~~~~~~~~~~~~~~~~~~~~~~~~~~~~~~~~~~~~~~~~~~~~~~~~~~~~~~~~~~~~\nonumber\\
\times\Big(1-\frac{\alpha}{2}
(1-\frac{\Omega_{\Lambda}}{c^{2}(1+\Omega_{k})}) \Big)+\frac{n\Omega_{\Lambda}}{c^{2}(1+\Omega_{k})}\Big] \Big\}\times
\Big\{(-1-2n+\frac{2}{3}n^{2}\omega)\times~~~~~~~~~~~~~~~~~~~~~~~~~~~~~~~~~~~~~~~~~~~~~~~~~~~~~~~~~~~~~~~~~~~~~~~~~~~~~~~~~~~~~~~~~~
~~~~~~~~~~~~~~~~~~~~~~~~~~~~~~~~~~~~~~~~~~~~~~~~~~~~~~~~~~~~~\nonumber\\\Big[-1-2n+\frac{2}{3}n^{2}\omega+3c^{2}H_{0}\Big(1-\frac{\alpha}{2}
(1-\frac{\Omega_{\Lambda}}{c^{2}(1+\Omega_{k})}) \Big)\times
\Big[\frac{\Omega_{M_{0}}(1+z)^{2n+3}}{1+\Omega_{k}+\frac{2}{3}n(3-n\omega)-\Omega_{\Lambda}} \Big]^{\frac{1}{2}}\Big]^{-1} \Big\}.~~~~~~~~~~~~~~~~~~~~~~~~~~~~~~~~~~~~~~~~~~~~~~~~~~~~~~~~~~~~~~~~~~~~~~~~~~~~~~~~~~~~~~~~~~~~~~~~~~~~~~~~~~~~~~~~~~~~~~~~~~~~~~~~~~~~~~~~~~~
\end{eqnarray}
For the present time, we consider $z\simeq0$,$\alpha\simeq0$,
 $c^{2}=1.1$ \cite{ma}, $n=0.005$ \cite{Lu}, $H_{0}=67.8$, $\Omega_{\Lambda}=0.7$, $\Omega_{M_{0}}=0.27$, $\Omega_{k}\simeq0$ and $b^{2}=0.02$ \cite{ma}.  By this set up, the conditions for acceleration $q<0$ and phantom
phase $\omega_{\Lambda}<-1$ results in $\omega<10000$ and  $-10000<\omega<10000$,
respectively. Therefore, for $-10000<\omega<10000$ we have both acceleration
and phantom phase.

At early time, using Eqs.(\ref{Omega1}) and $H=constant$,  we obtain $q\simeq-1$ and
\begin{eqnarray}
\omega_{\Lambda}=-1-\frac{2c^{2}H(1+\Omega_{k})}{\Omega_{\Lambda}}\Big[-\frac{\Omega_{k}}{1+\Omega_{k}}
\Big\{1-\frac{\alpha}{2}(1-
\frac{\Omega_{\Lambda}}{c^{2}(1+\Omega_{k})}) \Big\}
+\frac{n\Omega_{\Lambda}}{c^{2}(1+\Omega_{k})} \Big].\label{wa1}
\end{eqnarray} 
For early time, the condition $\omega_{\Lambda}<-1$ requires  $n>0$, so $n\simeq0.005$ \cite{Lu} is consistent with the phantom phase. Therefore, the PLECHDE model with apparent cut-off  and $-10000\lesssim\omega\lesssim10000$
supports the inflationary and accelerating universe with phantom phase at early and present universe, respectively.  

Using  $\omega_{\Lambda}$ and its time derivative corresponding to  early and present times, and using Eq.(\ref{vss}) with $b^{2}=0, 0.02, 0.04, 0.06$,  we get $v_{s}^{2}<0$ and $v_{s}^{2}>0$ respectively,
indicating that PLECHDE model has classical instability at early time and classical stability at present time.
\subsection{LECHDE model}
For interacting  logarithmic entropy-corrected holographic dark energy model
at present time, using Eqs.(\ref{Hz}), (\ref{dH}), and (\ref {Omega}) one can write
\begin{eqnarray}
\omega_{\Lambda}=\Big[-1-(n+\frac{\Omega_{k}}{(1+\Omega_{k})})[\frac{2c^{2}(1+\Omega_{k})}{3\Omega_{\Lambda}} +\frac{8\alpha \omega H_0^{2}\Omega_{M_{0}}(1+z)^{4n+3}(1+\Omega_{k})^{2}}{9\Omega_{\Lambda}(1+\Omega_{k}
+\frac{2n}{3}(3-n\omega)-\Omega_{\Lambda})}] +\frac{4\Omega_{k}}{3(1+\Omega_{k})}+~~~~~~~~~~~~~~~~~~~~~~~~~~
\\ \nonumber[\frac{-2\Omega_{k}+(2n+3+3b^2)(1+\Omega_{k}
-\frac{2n^{2}\omega}{3}+2n)}{2(-1+\frac{2}{3}n^2 \omega-2n)(1+\Omega_{k})}]\times [\frac{2c^{2}(1+\Omega_{k})}{3\Omega_{\Lambda}}+\frac{8\alpha \omega H_0^{2}\Omega_{M_{0}}(1+z)^{4n+3}(1+\Omega_{k})^{2}}{9\Omega_{\Lambda}(1+\Omega_{k}
+\frac{2n}{3}(3-n\omega)-\Omega_{\Lambda})}-\frac{4}{3}]~~~~~~~~~~~~~~~\\ \nonumber
-\frac{b^{2}}{\Omega_{\Lambda}}(1+\Omega_{k}-\frac{2}{3}n^2 \omega+2n)\Big]\times~~~~~~~~~~~~~~~~~~~~~~~~~~~~~~~~~~~~~~~~~~~~~~~~~~~~~~~~~~~~~~~~~~~~~~~~~~~~~~~~~~~~~~~~~~~~~~~~~~~~~~~~~~~~~~\\ \Big[1-(\frac{3\Omega_{\Lambda}}{2(-1+\frac{2}{3}n^{2}\omega-2n)(1+\Omega_{k})})\times \nonumber[\frac{2c^{2}(1+\Omega_{k})}{3\Omega_{\Lambda}}
+\frac{8\alpha \omega H_0^{2}\Omega_{M_{0}}(1+z)^{4n+3}(1+\Omega_{k})^{2}}{9\Omega_{\Lambda}(1+\Omega_{k}
+\frac{2n}{3}(3-n\omega)-\Omega_{\Lambda})}-\frac{4}{3}]\Big]_.^{-1}~~~~~~~~~~~~~~~~~~~~~~~~\label{wt}
\end{eqnarray}
For the present time, we consider $z\simeq0$, $c^{2}=1.1$ \cite{ma}, $n=0.005$ \cite{Lu}, $H_{0}=67.8$, $\Omega_{M_{0}}=0.27$, $\Omega_{k}\simeq0$ and $b^{2}=0.02$ \cite{ma}. By this set up, for $q<0$ and  $\omega_{\Lambda}<-1$ we obtain $-29032\lesssim\omega\lesssim34254$ and  $16908\lesssim\omega\lesssim38858$, respectively. Therefore, we may have acceleration with the phantom phase
for some regions of $\omega$.

At early time, using Eqs. (\ref{Omega}) and (\ref{apparent}) and $H=constant$, we obtain $q\simeq-1$ and
\begin{eqnarray}
\omega_{\Lambda}=-1-(n+\frac{\Omega_{k}}{(1+\Omega_{k})})[\frac{2c^{2}(1+\Omega_{k})}{3\Omega_{\Lambda}}+
\frac{8\alpha \omega H^{2}(1+z)^{2n}(1+\Omega_{k})^{2}}{9\Omega_{\Lambda}}]  +\frac{4\Omega_{k}}{3(1+\Omega_{k})},\label{wt1}
\end{eqnarray}
where $\alpha$ is dimensionless constant of order unity. 

Using the parameter values $z\simeq 10^4$, $\alpha\simeq 1$, $c^{2}\simeq1.1$ and $\Omega_{\Lambda}\simeq1$ , the condition $\omega_{\Lambda}<-1$ or $\omega_{\Lambda}>-1$ requires  $\omega\gtrsim-H^{-2}$ or $\omega\lesssim-H^{-2}$, respectively .

 Therefore,  LECHDE model with apparent cut-off  and $16908\lesssim\omega\lesssim38858$
supports the inflationary and accelerating universe with phantom phase at  present universe; however, at early universe we may have inflationary universe
with or without phantom phase corresponding to $\omega\gtrsim-H^{-2}$ or $\omega\lesssim-H^{-2}$, respectively.

 Using the same procedure as that of PLECHDE model, for LECHDE model at present time, we obtain that the squared
speed $v_{s}^{2}$ is  negative for $0.5<\Omega_{\Lambda}<1$ and  is  positive for $0<\Omega_{\Lambda}<0.5$. That is, we have classical instability  $v_{s}^{2}<0$, for $0.5<\Omega_{\Lambda}<1$ and the classical stability  $v_{s}^{2}>0$, for $0<\Omega_{\Lambda}<0.5$.

 At early time, we obtain  that the squared
speed is always negative and that there is
a classical instability.   

\section{Interacting  entropy-corrected  holographic dark energy model with event horizon cut-off}
The event horizon cut-off is considered as
\begin{equation}
L=a(t)r(t),\label{lr}
\end{equation}
and
\[
r(t)=\frac{\sin n (\sqrt{|k| y})}{\sqrt{|k|}}= \left\{%
\begin{array}{ll}
\sin y ~~~~~~~~k=1,\\
y ~~~~~~~~~~~~k=0,\\
\sinh y ~~~~~~k=-1,
\end{array}%
\right.
\]
where
\begin{equation}
y=\frac{R_{h}}{a(t)}=a(t)\int_{a(t)}^{\infty}{\frac{da(t)}{a(t)^{2}H}}.\label{yl}
\end{equation}
Here, $L$ and $R_{h}$ are the radius of the event horizon measured on the sphere of the horizon and the radial size of the event horizon, respectively \cite{Huang1}.

\subsection{PLECHDE model}

For power-law entropy-corrected holographic  dark energy model,
Using Eqs. (\ref{Hz}), (\ref{dH}), (\ref{Omega1}), (\ref{lr}) and (\ref{yl}) at the present time we can write
\begin{eqnarray}
\omega_{\Lambda}=-1-\frac{b^{2}}{\Omega_{\Lambda}}(1+\Omega_{k}-\frac{2}{3}n^{2}\omega+2n)
-\frac{2H_{0}}{\gamma_{c}}
\Big[\frac{\Omega_{M_{0}}(1+z)^{2n+3}}{1+\Omega_{k}+\frac{2}{3}n(3-n\omega)-\Omega_{\Lambda}} \Big]^{\frac{1}{2}}
\Big[-1+n\gamma_{c}+\sqrt{\frac{\Omega_{\Lambda}}{c^{2}\gamma_{c}}-\Omega_{k}} \Big]\label{wr}
\end{eqnarray}
where
\begin{equation}
\gamma_{c}=1-\frac{\beta}{3c^{2}}L^{2-\alpha}.\label{gammacc}
\end{equation}
Moreover, for PLECHDE at early time, using Eqs.  (\ref{Omega1}), (\ref{lr}), (\ref{yl}) and $H\simeq constant$, we obtain $q\simeq-1$ and
\begin{eqnarray}
\omega_{\Lambda}=-1-2Hc^{2}\Big[\Big(-1+\sqrt{\Omega_{\Lambda}-\Omega_{k}} \Big)\times
\Big(1-\frac{\alpha}{2}(1-\frac{1}{c^{2}}) \Big)+\frac{n}{c^{2}} \Big].\label{wr1}
\end{eqnarray}
For the present time, we consider $z\simeq0$,$\alpha\simeq0$,
 $c^{2}=1.1$ \cite{ma}, $n=0.005$ \cite{Lu}, $H_{0}=67.8$, $\Omega_{\Lambda}=0.7$, $\Omega_{M_{0}}=0.27$, $\Omega_{k}\simeq0$ and $b^{2}=0.02$ \cite{ma}.  By this set up, The condition for acceleration $q<0$  results in $-5000<\omega<0$
and no Brans-Dicke parameter is obtained for the phantom phase condition $\omega_{\Lambda}<-1$. Therefore, we have acceleration without the phantom phase.

Using  $\omega_{\Lambda}$ and its time derivative corresponding to  early and present times, and using Eq.(\ref{vss}) with $b^{2}=0, 0.02, 0.04, 0.06$,  we get the classical stability for the range $0<\Omega_{\Lambda}<0.2$ and the classical instability for the range $0.2<\Omega_{\Lambda}<1$ at present
time. At early time, we find that the squared
speed $v_{s}^{2}$ is  positive indicating  the classical stability
for early time.

\subsection{LECHDE model}

For logarithmic entropy-corrected holographic  dark energy model,
Using Eqs. (\ref{Hz}), (\ref{dH}), (\ref{Omega}), (\ref{lr}) and (\ref{yl}) at the present time we can write
\begin{eqnarray}
\omega_{\Lambda}=-1-n\Big[\frac{2}{3\gamma_{\alpha}}+\frac{8\alpha\omega(1+z)^{2n}}{9L^{2}c^{2}\gamma_{\alpha}}\Big]+
\Big(1-\sqrt{\frac{\Omega_{\Lambda}}{c^{2}\gamma_{\alpha}}-\Omega_{k}}\Big)\Big[-\frac{2}{3\gamma_{\alpha}}-\frac{8\alpha\omega(1+z)^{2n}}{9L^{2}c^{2}\gamma_{\alpha}}
+\frac{4}{3}\Big]\nonumber\\ -\frac{b^{2}}{\Omega_{\Lambda}}(1+\Omega_{k}-\frac{2}{3}n^2 \omega+2n),~~~~~~~~~~~~~~~~~~~~~~~~~~~~~~~~~~~~~~~~~~~~~~~~~~~~~~~~~~~~~~~~~~~~~~~\label{w1}
\end{eqnarray}
where
\begin{equation}
\gamma_{\alpha}=1+\frac{4\omega(1+z)^{2n}}{3L^{2}c^{2}}\Big[\alpha\ln(\frac{L^{2}}{4\omega(1+z)^{2n}}+\beta)\Big].
\label{gamac}
\end{equation}

Also, for LECHDE at the early time, using Eqs.  (\ref{Omega}), (\ref{lr}), (\ref{yl}) and $H=constant$, we obtain $q\simeq-1 $ and
\begin{eqnarray}
\omega_{\Lambda}=-1-n\Big[\frac{2c^{2}}{3}+\frac{8\alpha\omega(1+z)^{2n}H^{2}}{9}\Big]+
\Big(1-\sqrt{\Omega_{\Lambda}-\Omega_{k}}\Big)\Big[-\frac{2c^{2}}{3}-\frac{8\alpha\omega(1+z)^{2n}H^{2}}{9}
+\frac{4}{3}\Big].\label{w11}
\end{eqnarray}

For the present time, we consider $z\simeq0$,$\alpha\simeq0$,
 $c^{2}=1.1$ \cite{ma}, $n=0.005$ \cite{Lu}, $H_{0}=67.8$, $\Omega_{\Lambda}=0.7$, $\Omega_{M_{0}}=0.27$, $\Omega_{k}\simeq0$ and $b^{2}=0.02$ \cite{ma}.  By this set up, The conditions for acceleration $q<0$ and $\omega_{\Lambda}<-1$ results in $-51628\lesssim\omega\lesssim59086$ and  $\omega\lesssim-225706$,
respectively. Therefore, we have acceleration without the phantom phase.

At early time, Using the parameter values $z\simeq 10^4$, $\alpha\simeq 1$, $c^{2}\simeq1.1$ and $\Omega_{\Lambda}\simeq1$ , the condition $\omega_{\Lambda}<-1$ or $\omega_{\Lambda}>-1$ requires  $\omega\gtrsim-H^{-2}$ or $\omega\lesssim-H^{-2}$, respectively .

 Using the same procedure as that of PLECHDE model, for LECHDE model we obtain $v_{s}^{2}<0$ for present time and  $v_{s}^{2}>0$ for early  time which accounts for the classical instability at present time
and  the classical stability at early time
 .

\section{Interacting  entropy-corrected  holographic dark energy model with scalar Ricci cut-off}
We consider IR cut-off as $L=R^{-\frac{1}{2}}$ where $R$ is Ricci scalar curvature. The scalar Ricci cut-off is given by
\begin{equation}
R=6(\dot{H}+2H^{2}+\frac{k}{a^{2}}).\label{LRicci}
\end{equation}
Here $\dot{H}$ is the derivative of the hubble parameter with respect to the cosmic time $t$. Using Eqs. (\ref{frid}) and (\ref{scale}), we get \cite{Pasqua}
\begin{equation}
H^{2}+\frac{k}{a^{2}}=\frac{4\omega}{3\phi^{2}}(\rho_{\Lambda}+\rho_{M})+2nH^{2}(-1+\frac{n\omega}{3}).\label{friddd}
\end{equation}
Now, using  Eqs. (\ref{LRicci}) and (\ref{friddd}), we can write
\begin{equation}
R=6\Big\{\dot{H}+H^{2}+\frac{4\omega}{3\phi^{2}}(\rho_{\Lambda}+\rho_{M})+2nH^{2}(-1+\frac{n\omega}{3}) \Big\}.\label{RRR}
\end{equation}
Differentiating Eq. (\ref{friddd}) with respect to the cosmic time $t$ and using Eqs. (\ref{dot rho}), (\ref{QM}) and (\ref{friddd}), we can derive
\begin{equation}
\dot{H}+H^{2}=\frac{\frac{4\omega}{3\phi^{2}}\Big[\rho_{\Lambda}(\frac{3}{2}\omega_{\Lambda}+\frac{1}{2}+n)+(n+\frac{1}{2})\rho_{M} \Big]}{-1+2n(\frac{n\omega}{3}-1)}.\label{Hdott}
\end{equation}
Inserting Eg. (\ref{Hdott}) in Eq. (\ref{RRR}), we obtain
\begin{eqnarray}
\omega_{\Lambda}=\Big[\frac{\phi^{2}(2n^{2}\omega-6n-3)}{\rho_{\Lambda}}\Big]\Big[\frac{R}{36\omega}-\frac{nH^{2}(n\omega-3)}{9\omega} \Big]
-\frac{(1+\Omega_{k}-\frac{2}{3}n^{2}\omega+2n)(4n^{2}\omega-6n-3)}{9\Omega_{\Lambda}}.~~~~~~~\label{omegar}
\end{eqnarray}
Moreover, using Eqs. (\ref{Plambda}), (\ref{scale}), (\ref{Lambda}), (\ref{deceleration}) and $P_{\Lambda}=\rho_{\Lambda}\omega_{\Lambda}$ we find
\begin{eqnarray}
q=\frac{1}{2(n+1)}\times\Big[3\Omega_{\Lambda}\omega_{\Lambda}+(2n+1)^{2}+2n(n\omega-1)+
\Omega_{k} \Big].~~~~~
\label{qRicci}
\end{eqnarray}
\subsection{PLECHDE model}
 Using Eq.(\ref{rho1}) and $L=R^{-\frac{1}{2}}$, we find
\begin{equation}
\rho_{\Lambda}=\frac{3c^{2}\phi^{2}R}{4\omega}\gamma_{\mu},\label{rhor}
\end{equation}
where
\begin{equation}
\gamma_{\mu}=1-\frac{\beta}{3c^{2}}R^{\frac{\alpha}{2}-1}.
\end{equation}
For early time, using Eqs. (\ref{Omega1}), (\ref{LRicci}) and $H=constant$, we can obtain $q\simeq-1$ and 
\begin{eqnarray}
\omega_{\Lambda}=-1+\frac{12c^{2}H}{\Omega_{\Lambda}}\Big[-1+\frac{2}{3}n^{2}\omega-2n+\Omega_{\Lambda} \Big]\times\Big\{1-\frac{\alpha}{2}(1-\frac{H^{2}\Omega_{\Lambda}}{c^{2}R}) \Big\}-2nH.
\label{wricci1}
\end{eqnarray}
Here $\alpha$  is dimensionless constants of order unity. For  early time, we consider $c^{2}=1.1$, $\Omega_{k}\simeq0$ \cite{ma} and   assume $\alpha>2$ \cite{Sheykhi}.  In order to have $\omega_{\Lambda}<-1$, it turns out that $n>0$ for any value of $\omega$.

For $0<R<1$ at the early time, taking time derivative of $\omega_{\Lambda}$, and using Eqs. (\ref{vss}) and (\ref{wricci1}),  one finds that the squared speed is always negative for early time. This means that we  have a classical instability.

\subsection{LECHDE model}
 Using Eq.(\ref{rho}) and $L^{2}=R^{-1}$, one can obtain
\begin{equation}
\rho_{\Lambda}=\frac{3c^{2}\phi^{2}R}{4\omega}\gamma_{\phi},\label{rhor1}
\end{equation}
where
\begin{equation}
\gamma_{\phi}=1+\frac{4\omega R}{3c^{2}\phi^{2}}\Big[\alpha \ln(\frac{\phi^{2}}{4\omega R})+\beta \Big].
\end{equation}
For the early time, using Eqs.(\ref {LRicci}) and (\ref {Omega}) , $H\simeq
constant$,  we obtain $q\simeq-1$ and 
\begin{eqnarray}
\omega_{\Lambda}=-1-n\Big[\frac{8c^{2}}{\Omega_{\Lambda}}+\frac{128\alpha\omega(1+z)^{2n}H^{2}}{\Omega_{\Lambda}}\Big]\nonumber\\
\label{ome1}
\end{eqnarray}
Where $\alpha$  is dimensionless constants of order unity. For the early time, we consider $c^{2}=1.1$, $\Omega_{k}\simeq0$ \textcolor[rgb]{1,0,0.501961}{}\cite{ma}.  In order to have $\omega_{\Lambda}<-1$,
 we obtain the range $-10^{-98}<\omega<-10^{-42}$ and $n>-0.1$.  
For $0<R<1$ at the early time, 
 taking time derivative of Eq.(\ref{ome1}) and using Eqs.(\ref{vss}) and (\ref{ome1}),  one finds that the squared speed is always negative for early time. This means that we  have a classical instability.

For both PLECHDE and LECHDE models at the present time, we consider $\alpha=\beta=0$,  $\gamma_{\mu}=1$, $\gamma_{\phi}=1$.
Using Eqs. (\ref{Hz}), (\ref{omegar}), (\ref{rhor}) and (\ref{rhor1})  we obtain  \begin{eqnarray}
\omega_{\Lambda}=\Big[\frac{4(2n^{2}\omega-6n-3)}{3c^{2}} \Big]\times
\Big[\frac{1}{36}-\frac{nH_{0}^{2}(n\omega-3)\Omega_{M_{0}}(1+z)^{2n+3}}{9R\Big(1+\Omega_{k}+\frac{2}{3}n(3-n\omega)-\Omega_{\Lambda}\Big)} \Big]
~~~~~~~~~\nonumber\\
-\frac{(1+\Omega_{k}-\frac{2}{3}n^{2}\omega+2n)(4n^{2}\omega-6n-3)}{9\Omega_{\Lambda}}.\label{wricci}
\end{eqnarray}
Now, for both PLECHDE and LECHDE models, we consider $c^{2}=1.1$ \cite{ma}, $n=0.005$ \cite{Lu}, $H_{0}=67.8$, $\Omega_{M_{0}}=0.27$, $\Omega_{k}\simeq0$  \cite{ma} and $0<R<1$. By this set up, for $q<0$  and $\omega_{\Lambda}<-1$, we obtain $20000<\omega<45000$ and $-100000<\omega<-10000$, respectively. Therefore, we have acceleration without the phantom phase.

Taking time derivative of Eq.(\ref{wricci}) and using Eqs.(\ref{vss}) and (\ref{wricci}) for $b^{2}=0, 0.02, 0.04, 0.06$ and $0<R<1$ at the present
time,
one finds that the squared speed is always positive for present time. This means that we  have a classical stability.

\section{Interacting  entropy-corrected  holographic dark energy model with Granda-Oliveros cut-off}
To avoid the causality problem, Granda and Oliveros suggested a new cut-off for holographic dark energy model so called new holographic dark energy as
\cite{Oliveros}
\begin{equation}
L=(\tilde{\alpha}H^{2}+\tilde{\beta}\dot{H})^{-\frac{1}{2}},
\end{equation}
where $\tilde{\alpha}$ and $\tilde{\beta}$ are constant. By inserting Eq. (\ref{scale}) in Eq. (\ref{Plambda}) and using $P_{\Lambda}=\rho_{\Lambda}\omega_{\Lambda}$, we obtain
\begin{eqnarray}
\rho_{\Lambda}=-\frac{\phi^{2}H^{2}}{4\omega\omega_{\Lambda}}\Big\{3+\Omega_{k}+4n+2n^{2}(2+\omega)
+\frac{\dot{H}}{H^{2}}(2n+2) \Big\}. \label{rhoGo}
\end{eqnarray}
moreover,  By inserting Eq. (\ref{dH}) in Eq. (\ref{rhoGo}), we find
\begin{eqnarray}
\omega_{\Lambda}=\Big[3+\Omega_{k} +4n+2n^{2}(2+\omega)+(n+1)\times
\Big(\frac{(2n+3+3b^{2})(1+\Omega_{k}-\frac{2}{3}n^{2}\omega+2n)-2\Omega_{k}}{-1-2n+\frac{2}{3}n^{2}\omega}\Big)\Big]\times~~~~~~~~~~~~\nonumber\\
\Big[-\frac{4\omega \rho_{\Lambda}}{\phi^{2}H^{2}}
-\frac{3(n+1)\Omega_{\Lambda}}{-1-2n+\frac{2}{3}n^{2}\omega} \Big]^{-1}.~~~~~~~~~~~~~~~~~~~~~~~~
\end{eqnarray}
For PLECHDE and LECHDE models, using Eq. (\ref{rho1}) and Eq.(\ref{rho}), we find
respectively\begin{equation}
\rho_{\Lambda}=\frac{3c^{2}\phi^{2}}{4\omega L^{2}}\gamma_{c},\label{ggoo}
\end{equation}
 and
\begin{equation}
\rho_{\Lambda}=\frac{3c^{2}\phi^{2}}{4\omega L^{2}}\gamma_{\alpha}, \label{rGO}
\end{equation}
where $L$  is  the Granda-Oliveros cut off.  For both PLECHDE and LECHDE models,  considering $\alpha=\beta=0,$ and using $\omega_{\Lambda}$, (\ref{Hz}),  (\ref{ggoo}) and
(\ref{rGO}) at  present time we can find
\begin{eqnarray}
\omega_{\Lambda}=\Big\{3+\Omega_{k}+4n+2n^{2}(2+\omega)+(n+1)\times
\Big[\frac{(2n+3+3b^{2})(1+\Omega_{k}-\frac{2}{3}n^{2}\omega+2n)-2\Omega_{k}}{-1-2n+\frac{2}{3}n^{2}\omega} \Big] \Big\}\times~~~~~~~~~~~~~\nonumber\\
\Big\{-\frac{3c^{2}\Big(1+\Omega_{k}+\frac{2}{3}n(3-n\omega)-\Omega_{\Lambda}\Big)}{L^{2}H_{0}^{2}\Omega_{M_{0}}(1+z)^{2n+3}}
-\frac{3(n+1)\Omega_{\Lambda}}{-1+\frac{2}{3}n^{2}\omega-2n}
\Big\}^{-1}, ~~~~~~~~~~~~~~~~~~~~~~~~
\end{eqnarray}
where $q$ is given by Eq. (\ref{qRicci}). For the present time,  we consider $c^{2}=1.1$ \cite{ma}, $n=0.005$ \cite{Lu}, $H_{0}=67.8$, $\Omega_{M_{0}}=0.27$, $\Omega_{k}\simeq0$ and $b^{2}=0.02$  \cite{ma}, $z=0$ and $0<L<1$ . For both LECHDE and   PLECHDE models, the demand for $q<0$ and $\omega_{\Lambda}<-1$ results in $10000<\omega<12000$ and $11100<\omega<12000$, respectively.
Hence, we may have acceleration with phantom phase for $11100<\omega<12000$.

For the early time, using Eqs. (\ref{rhoGo}), (\ref{ggoo}), (\ref{rGO}) and $H=constant$,  we can obtain $q\simeq-1$ and
\begin{equation}
\omega_{\Lambda}=-\frac{3+\Omega_{k}+4n+2n^{2}(2+\omega)}{3\Omega_{\Lambda}}.\label{omegaaa1}
\end{equation}
 Considering $c^{2}=1.1$ and $\Omega_{k}\simeq0$ \cite{ma} we find that the
demand for $\omega_{\Lambda}<-1$, results in $n>0$.

For present time, taking time derivative of $\omega_{\Lambda}$ and using Eqs.(\ref{vss}) and $\omega_{\Lambda}$ for $b^{2}=0, 0.02, 0.04, 0.06$ and $0<L<1$, we obtain the classical instability. For both PLECHDE model and LECHDE model, one finds that the squared speed is negative. This means that we have a classical instability.

For early time, taking the time derivative of $\omega_{\Lambda}$, and using Eqs.(\ref{vss}) and $\omega_{\Lambda}$,  we find that for both PLECHDE and LECHDE models  the squared speed is always negative. This means that we have a classical instability.


\hspace{3mm}{\small {\bf  Table 1.}} {\small
  Brans-Dicke parameter versus the phantom phase $\omega_{\Lambda}<-1$ at  present time.}\\
    \begin{tabular}{l l l l l p{0.15mm} }
    \hline\hline
  \vspace{0.5mm}
{\footnotesize  Cut-Off  } & {\footnotesize  \ ~~~~~~~~~~~~~~~~~~~~      PLECHDE }  &  
{\footnotesize  \  ~~~~~~~~~~~~~~~~~~~~~~~~~~~~~~~~~ LECHDE }&  \\\hline

\vspace{2mm}

{\footnotesize Hubble}& {\footnotesize~~~~~~~~~~~~~~ $-10000<\omega<10000$}&  
{\footnotesize~~~~~~~~~~~~~~~~~~~~~~~~~~~~~~~~~ $   \omega\lesssim-2860$}&\\\hline

\vspace{2mm}

{\footnotesize Apparent}& {\footnotesize ~~~~~~~~~~~~~~~$-10000<\omega<10000$} & 
{\footnotesize~~~~~~~~~~~~~~~~~~~~~~~~~~~~ $16908\lesssim\omega\lesssim38858$}&\\\hline

\vspace{2mm}

{\footnotesize {Event Horizon}}& {\footnotesize~~~~~~~~~~~~~~~~~~~~~~~~ $---$} &
 {\footnotesize ~~~~~~~~~~~~~~~~~~~~~~~~~~~~~~~~~$ \omega\lesssim-225706$}&{\footnotesize $$}\\\hline

\vspace{2mm}

{\footnotesize Scalar Ricci}&  {\footnotesize~~~~~~~~~~~~~~ $-100000<\omega<-10000$} &{\footnotesize~~~~~~~~~~~~~~~~~~~~~~~ $-100000<\omega<-10000$}&  {\footnotesize $$}\\\hline

\vspace{2mm}

{\footnotesize Granda-Oliveros}& {\footnotesize~~~~~~~~~~~~~~~~~ $11100<\omega<12000$} &  {\footnotesize~~~~~~~~~~~~~~~~~~~~~~~~~~~ $11100<\omega<12000$} \\\hline

\vspace{12mm}
 \end{tabular}

\hspace{3mm}{\small {\bf Table 2.}} {\small
  Brans-Dicke parameter versus the acceleration phase $q<0$ at  present time.}\\
    \begin{tabular}{l l l l l p{0.15mm} }
    \hline\hline
  
{\footnotesize  Cut-Off  } & {\footnotesize  \ ~~~~~~~~~~~~~~~~~~~~      PLECHDE }  &  
{\footnotesize  \  ~~~~~~~~~~~~~~~~~~~~~~~~~~~~~~~~~ LECHDE }&  \\\hline

\vspace{2mm}

{\footnotesize Hubble}& {\footnotesize~~~~~~~~~~~~~~ $-10000<\omega<10000$}&  
{\footnotesize~~~~~~~~~~~~~~~~~~~~~~~~~~~~~~~~~ $\omega\lesssim8383$}&\\\hline

\vspace{2mm}

{\footnotesize Apparent}& {\footnotesize ~~~~~~~~~~~~~~~$-10000<\omega<10000$} & 
{\footnotesize~~~~~~~~~~~~~~~~~~~~~~~~~~~~ $-29032\lesssim\omega\lesssim34254$}&\\\hline

\vspace{2mm}

{\footnotesize {Event Horizon}}& {\footnotesize~~~~~~~~~~~~~~~~ $-5000<\omega<0$} &
 {\footnotesize ~~~~~~~~~~~~~~~~~~~~~~~~~~~~~$-51628\lesssim\omega\lesssim59086$}&{\footnotesize $$}\\\hline

\vspace{2mm}

{\footnotesize Scalar Ricci}&  {\footnotesize~~~~~~~~~~~~~~~~ $20000<\omega<45000$} &{\footnotesize~~~~~~~~~~~~~~~~~~~~~~~~~~~~~~~$20000<\omega<45000$ }&  {\footnotesize $$}\\\hline

\vspace{2mm}

{\footnotesize Granda-Oliveros}& {\footnotesize~~~~~~~~~~~~~~~~ $10000<\omega<12000$} &  {\footnotesize~~~~~~~~~~~~~~~~~~~~~~~~~~~~~~ $10000<\omega<12000$} \\\hline

\vspace{12mm}
 \end{tabular}

\hspace{3mm}{\small {\bf Table 3.}} {\small
  Brans-Dicke parameter versus the phantom phase $\omega_{\Lambda}<-1$ at  early time.}\\
    \begin{tabular}{l l l l l p{0.15mm} }
    \hline\hline
  \vspace{0.5mm}
{\footnotesize  Cut-Off  } & {\footnotesize  \ ~~~~~~~~~~~~~~~~~~~~      PLECHDE }  &  
{\footnotesize  \  ~~~~~~~~~~~~~~~~~~~~~~~~~~~~~~ LECHDE }&  \\\hline

\vspace{2mm}

{\footnotesize Hubble}& {\footnotesize~~~~~~~~~~~~~~ $n>0,-\infty<\omega<\infty$}&  
{\footnotesize~~~~~~~~~~~~~~~~~~~~~~~~~~~~~~ $\omega>-H^{-2}$}&\\\hline

\vspace{2mm}

{\footnotesize Apparent}& {\footnotesize ~~~~~~~~~~~~~~ $n>0,-\infty<\omega<\infty$} & 
{\footnotesize~~~~~~~~~~~~~~~~~~~~~~~~~~~~~~ $\omega>-H^{-2}$}&\\\hline

\vspace{2mm}

{\footnotesize {Event Horizon}}& {\footnotesize~~~~~~~~~~~~~~ $n>0,-\infty<\omega<\infty$} & 
{\footnotesize~~~~~~~~~~~~~~~~~~~~~~~~~~~~~~ $\omega>-H^{-2}$}&{\footnotesize $$}\\\hline

\vspace{2mm}

{\footnotesize Scalar Ricci}&  {\footnotesize~~~~~~~~~~~~~~ $n>0,-\infty<\omega<\infty$} &{\footnotesize~~~~~~~~~~~~~~ $-10^{-98}<\omega<-10^{-42},
n>-0.1$}&  {\footnotesize $$}\\\hline

\vspace{2mm}

{\footnotesize Granda-Oliveros}& {\footnotesize~~~~~~~~~~~~~~~~~~~~ $n>0,~\omega>0$} &  {\footnotesize~~~~~~~~~~~~~~~~~~~~~~~~~~~~~ $n>0,~\omega>0$} \\\hline

\vspace{12mm}
 \end{tabular}

\hspace{3mm}{\small {\bf Table 4.}} {\small
  Brans-Dicke parameter versus the acceleration phase $q<0$ at  early time.}\\
    \begin{tabular}{l l l l l p{0.15mm} }
    \hline\hline
  \vspace{0.5mm}
{\footnotesize  Cut-Off  } & {\footnotesize  \ ~~~~~~~~~~~~~~~~~~~~~~~~~~~~~~~~      PLECHDE }  &  
{\footnotesize  \  ~~~~~~~~~~~~~~~~~~~~~~~~~~~~~~ LECHDE }&  \\\hline

\vspace{2mm}

{\footnotesize Hubble}& {\footnotesize~~~~~~~~~~~~~~~~~~~~~~~~~ $\omega=-\frac{3(3H-2n-4)}{n(2n+3)}, n\neq0,-\frac{3}{2}$}&  
{\footnotesize~~~~~~~~~~~~~~~~~~ $\omega=\frac{6n+8.7}{2n^{2}+3n+4H^2(10001)^{2n}},-\infty<n<\infty$}

&\\\hline

\vspace{2mm}

{\footnotesize Apparent}& {\footnotesize ~~~~~~~~~~~~~~~~~~~~~~~~~ $\omega=-\frac{3(3H-2n-4)}{n(2n+3)}, n\neq0,-\frac{3}{2}$}&  
{\footnotesize~~~~~~~~~~~~~~~~~~ $\omega=\frac{6n+8.7}{2n^{2}+3n+4H^2(10001)^{2n}},-\infty<n<\infty$}&\\\hline

\vspace{2mm}

{\footnotesize {Event Horizon}}& {\footnotesize~~~~~~~~~~~~~~~~~~~~~~~~~ $\omega=-\frac{3(3H-2n-4)}{n(2n+3)}, n\neq0,-\frac{3}{2}$}&  
{\footnotesize~~~~~~~~~~~~~~~~~~ $\omega=\frac{6n+8.7}{2n^{2}+3n+4H^2(10001)^{2n}},-\infty<n<\infty$}&{\footnotesize $$}\\\hline

\vspace{2mm}

{\footnotesize Scalar Ricci}&  {\footnotesize~~~~~~~~~~~~~~~~~~~~~~~~~ $\omega=\frac{33.75H^3-10.5H-1.25n-1.25}{n(11.25H^3-4.125H+0.625)},n\neq0$}&  
{\footnotesize~~~~~~~~~~~~~~~~~~ $\omega=\frac{6n-33.6}{-3n+64H^2(10001)^{2n}},~~-\infty<n<\infty$ }&  {\footnotesize $$}\\\hline

\vspace{2mm}

{\footnotesize Granda-Oliveros}& {\footnotesize~~~~~~~~~~~~~~~~~~~~~~~~~~~~~~~~~~~~~~ $n=0$}&  
{\footnotesize~~~~~~~~~~~~~~~~~~~~~~~~~~~~~~~~~~ $n=0$} \\\hline
\vspace{2mm}
 \end{tabular}
 
\hspace{3mm}{\small {\bf Table 5.}} {\small
Classical stability or instability of models at  present time.}\\
    \begin{tabular}{l l l l l p{0.15mm} }
    \hline\hline
  \vspace{0.5mm}
{\footnotesize  Cut-Off  } & {\footnotesize ~~~~~~~~~~~~~~~ PLECHDE }  &  
{\footnotesize ~~~~~~~~~~~~~~~~~ LECHDE }&  \\\hline

\vspace{2mm}

{\footnotesize Hubble}& {\footnotesize ~~~~~~~~~~~~~~~~~~Stability}&  
 {\footnotesize ~~~~~~~~~~~~~~~~~~Instability}&\\\hline

\vspace{2mm}

{\footnotesize Apparent}& {\footnotesize ~~~~~~~~~~~~~~~~~~Stability} & 
{\footnotesize ~~~~~~~~~~Instability$(0 < \Omega_{\Lambda}<0.5)$}& &\\{\footnotesize $$}&{\footnotesize $$}&  {\footnotesize 
~~~~~~~~~~Stability$~~(0.5<\Omega_{\Lambda}<1)$}
\\\hline

\vspace{3mm}

{\footnotesize Event Horizon}& {\footnotesize ~~~~~~~~~~Stability$(0<\Omega_{\Lambda}<0.2)$}&{\footnotesize ~~~~~~~~~~~~~~~~~~Instability}\\{\footnotesize $$}& {\footnotesize ~~~~~~~~~~Instability$(0.2<\Omega_{\Lambda}<1)$}  
 \\\hline

\vspace{2mm}

{\footnotesize Scalar Ricci}&  {\footnotesize ~~~~~~~~~~~~~~~~~~Stability} &{\footnotesize ~~~~~~~~~~~~~~~~~~Stability}\\\hline

\vspace{2mm}

{\footnotesize Granda-Oliveros}& {\footnotesize~~~~~~~~~~~~~~~~ Instability} &  {\footnotesize ~~~~~~~~~~~~~~~~~ Instability} \\\hline

\vspace{5mm}
 \end{tabular}
 

\hspace{3mm}{\small {\bf Table 6.}} {\small
Classical stability or instability of models at  early time.}\\
    \begin{tabular}{l l l l l p{0.15mm} }
    \hline\hline
  \vspace{0.5mm}
{\footnotesize  Cut-Off  } & {\footnotesize ~~~~~~~~~~~~~~~~~~~~~ PLECHDE }  &  
{\footnotesize ~~~~~~~~~~~~~~~~~~~~~~~ LECHDE }&  \\\hline

\vspace{2mm}

{\footnotesize Hubble}& {\footnotesize ~~~~~~~~~~~~~~~~~~~~~~~Instability}&  
 {\footnotesize ~~~~~~~~~~~~~~~~~~~~~~~~Instability}&\\\hline

\vspace{2mm}

{\footnotesize Apparent}& {\footnotesize ~~~~~~~~~~~~~~~~~~~~~~~Instability} & 
{\footnotesize ~~~~~~~~~~~~~~~~~~~~~~~~Instability}
\\\hline

\vspace{3mm}

{\footnotesize Event Horizon}& {\footnotesize ~~~~~~~~~~~~~~~~~~~~~~~Stability} & 
{\footnotesize ~~~~~~~~~~~~~~~~~~~~~~~~Stability}
 \\\hline

\vspace{2mm}

{\footnotesize Scalar Ricci}&  {\footnotesize ~~~~~~~~~~~~~~~~~~~~~~~Instability} & 
{\footnotesize ~~~~~~~~~~~~~~~~~~~~~~~Instability}\\\hline

\vspace{2mm}

{\footnotesize Granda-Oliveros}& {\footnotesize~~~~~~~~~~~~~~~~~~~~~~ Instability} &  {\footnotesize ~~~~~~~~~~~~~~~~~~~~~~ Instability} \\\hline

 \end{tabular}


\section{Concluding remarks}\label{Con}

 We have studied the interacting power-law and logarithmic entropy-corrected holographic dark energy models with different cut-offs in the framework of Brans-Dicke cosmology. For comparison between the two entropy corrected models
at early and present time universes, we have obtained the Brans-Dicke parameter  versus the conditions for the acceleration and phantom phases.  Moreover, using the squared sound speed, the classical stability
or instability of the interacting power-law and logarithmic entropy-corrected holographic dark energy models with different cut-offs 
is determined. This study, shows which entropy corrected model can exhibit acceleration with or without the phantom phase, and which entropy corrected model is stable or unstable, at early and present time.


\end{document}